# QUADRUPOLE MOMENTS OF HIGH-SPIN ISOMERS: TEST OF THE TILTED-AXIS CRANKING MODEL

D.L. BALABANSKI[1,2], K. VYVEY[1], G. NEYENS[1],
D. ALMEHED[3], P. BLAHA[4], D. BORREMANS[1], S. CHMEL[5],
N. COULIER[1], R. COUSSEMENT[1], W. DE CLERCQ[1],
S. FRAUENDORF[3,6], G. GEORGIEV[1,2], H. HUBEL[5],
A. LÉPINE-SZILY[1,7], M. MINEVA[8], N. NENOFF[5],
S. PANCHOLI[5], D. ROSSBACH[5], R. SCHWENGNER[3],
S. TEUGHELS[1], P.M. WALKER[9]

[1] IKS, University of Leuven, Celestijnenlaan 200 D, B-3001 Leuven, Belgium
[2] Faculty of Physics, University of Sofia, BG-1164 Sofia, Bulgaria
[3] IKHP, FZ-Rossendorf, D-01314 Dresden, Germany
[4] IPTC, Technological University of Vienna, A-1060 Vienna, Austria
[5] ISKP, University of Bonn, Nussallee 14-16, D-53115 Bonn, Germany
[6] Department of Physics, University of Notre Dame, Indiana 46556, USA
[7] Instituto de Fisica, Universidade de Sao Paulo, 05389 Sao Paulo, Brazil
[8] Department of Physics, Lund University, S-221 00 Lund, Sweden
[9] Department of Physics, University of Surrey, Guildford GU2 7XH, UK

We report the results of recent measurements of the spectroscopic quadrupole moments of high-spin isomers. For the $K^\pi = \frac{35}{2}^-$ five-quasiparticle isomer in $^{179}$W we measured $Q_s = 4.00(^{+0.83}_{-1.06})$ $e$b. It corresponds to a smaller deformation compared to the ground states of the W isotopes and is in disagreement with the current theoretical predictions. We also measured the quadrupole moment of the $I^\pi = 11^-$ isomer in $^{196}$Pb, $Q_s = (-)3.41(66)$ $e$b. It has the same proton $(s_{1/2}^{-2} h_{9/2} i_{13/2})$ configuration as the one suggested for the $I^\pi = 16^{(-)}$ magnetic bandhead which allows to deduce the quadrupole moment of the $16^-$ state as $Q_s = -0.316(97)$ $e$b. This small value proves the near sphericity of the bandhead.

PACS numbers: 21.10.Ky, 21.10.Re, 23.20.Lv, 27.60.+q, 21.60.Ev

## 1. Introduction

Several measurements of spectroscopic quadrupole moments of high-spin isomers were performed at the CYCLONE facility at Louvaine-la-Neuve, using the <u>L</u>evel <u>M</u>ixing <u>S</u>pectroscopy technique (LEMS) [1].

We measured the quadrupole moment of the five-quasiparticle isomer in $^{179}$W, addressing the question: is the nuclear deformation of the ground state and of the high-seniority multiquasiparticle excitations the same?

Next, we studied the isomers in $N = 82$ nucleus $^{196}$Pb, for which several bands with enhanced $M1$ transitions have been established [2]. They are built on states which result from the coupling of high-spin particle to high-spin hole excitations. In the Pb nuclei, high-spin particle states are formed by proton excitations into the $\pi h_{9/2}$ and $\pi i_{13/2}$ orbitals and the high-spin hole states result from neutron excitations in the $\nu i_{13/2}$ shell. The specific coupling of the spin vectors of these two excitations causes a magnetic dipole moment with a large component perpendicular to the total spin. Its rotation around the spin axis gives rise to enhanced $M1$ radiation. Our study addresses the question: what is the deformation of the magnetic bandhead?

These experiments are related to the tilted-axis cranking (TAC) model, since both excitations, the high-$K$ states and the magnetic rotation are described within this approach [3, 4]. In the former case the quenching of pairing is studied. The usual approach is to deduce the moments of inertia of the bands built on the high-$K$ states and compare them to that of the ground-state (fully paired) rotational band [5]. Yet, the moment of inertia depends on both, the deformation and the pairing, which requires that the deformation is determined experimentally. In the case of magnetic rotation, the measured weak $B(E2)$ transition probabilities are an indication for small deformation [6]. However, the measurement of the spectroscopic quadrupole moment will provide the stringent test.

## 2. Experimental details

The LEMS set-up consists of a split-coil 4.4 T superconducting magnet, a target holder, allowing precise temperature control in the interval $4 - 600$ K, and 4 Ge detectors, positioned at $0°$ and $90°$, which monitor the target through the holes of the magnet. The magnetic field is oriented along the beam axis. The anisotropy of the $\gamma$-radiation is measured as a function of the external magnetic field. At small magnetic fields the initial anisotropy is perturbed due to the interaction of the quadrupole moment of the isomer of interest, $Q_s$, with the electric-field gradient (EFG) of the LEMS host, $V_{zz}$. It is restored at high fields. The transition between the two regimes depends on the ratio of the quadrupole frequency $\nu_Q = \frac{e}{h} \cdot Q_s \cdot V_{zz}$ and the magnetic moment of the isomer [1].

### 2.1. The case of $^{179}$W

The $^{170}$Er($^{13}$C,4n) reaction at 63 MeV was used to populate high-spin states in $^{179}$W. The target was 500 $\mu$g/cm$^2$ self-supporting, enriched $^{170}$Er, which allowed an in-beam implantation of the $^{179}$W recoils into a Tl foil at $T = 473(1)$ K, serving as a LEMS host. The obtained LEMS curve is presented in Fig. 2.1. A quadrupole frequency $\nu_Q = 53(8)$ MHz was deduced. A more complete report of this experiment was published elsewhere [9].

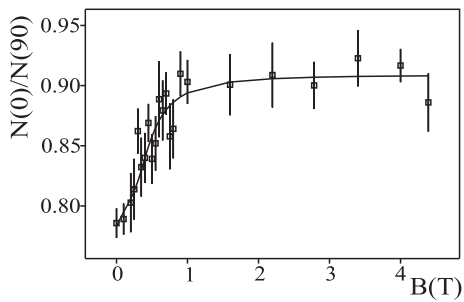

Fig. 1. LEMS curve for the $I = K = \frac{35}{2}$ keV isomer in $^{179}$W. The recoiling nuclei were implanted into a Tl polycrystalline foil at a temperature of 473(1) K.

In order to determine the EFG of WTl we performed band-structure calculations based on density functional theory using the full-potential linearized augmented plane wave (LAPW) method as implemented in the WIEN97 package [8]. A value of $V_{zz}(\text{WTl}) = 2.54 \cdot 10^{21}$ V/m$^2$ at 0 K was obtained. The EFGs in Tl decrease strongly with temperature. The temperature dependence of the EFGs in non-cubic metals follows the $T^{3/2}$ law: $V_{zz}(T) = V_{zz}(0) \cdot [1 - bT^{3/2}]$ [7]. In a dedicated experiment the temperature dependence factor $b = 7.6(^{+0.2}_{-0.4}) \cdot 10^{-5}$ K$^{-3/2}$ was derived, resulting in a value $V_{zz}(\text{WTl}) = 0.55(^{+0.12}_{-0.08}) \cdot 10^{21}$ V/m$^2$ at 473 K [10]. Thus, the spectroscopic quadrupole moment of the $K^\pi = \frac{35}{2}^-$ isomer in $^{179}$W is found to be $Q_s = 4.00(^{+0.83}_{-1.06})$ eb. This value is smaller than the ground state quadrupole moments of the $_{74}$W nuclei, which were derived from the reduced transition probabilities [11] and is about $2\sigma$ off the current theoretical estimates.

A critical reader might question the reliability of this result, as it relies on a calculated EFG for WTl, as well as on the assumption for a $T^{3/2}$ temperature dependence of $V_{zz}$. However, in a parallel experiment Ionescu-Bujor *et al.* measured the quadrupole moment of the $K^\pi = 14^+$ isomer in $^{176}$W [12]. The experiment was performed under similar conditions as our measurement (in-beam implantation of W in Tl at 464 K) and yields a quadrupole frequency $\nu_Q = 92(10)$ MHz for the isomer. This frequency corresponds to a ratio of the quadrupole moments of the two isomers:

$$\frac{Q_s(^{176}W)}{Q_s(^{179}W)} = 1.6 \pm 0.3, \tag{1}$$

and to value for the quadrupole moment of the $14^+$ isomer, which is close to the ground-state values. This is in disagreement with the current theoretical predictions and leads to the logical question: do we understand the deformations of the high-$K$ excitations?

## 2.2. The case of $^{196}Pb$

High-spin isomers in $^{196}$Pb (as well as in $^{194}$Pb) have been populated in the $^{nat}$Re($^{14}$N,5n) reaction. The 50 $\mu$m $^{nat}$Re foil served as a target and a LEMS host. Quadrupole frequencies of $\nu_Q = 38(3)$ MHz and $\nu_Q = 199(32)$ MHz were measured for the $I^\pi = 12^+$ and the $I^\pi = 11^-$ isomers, respectively. This study is discussed in more detail elsewhere [10]. The LEMS curve for the $I^\pi = 11^-$ isomer is displayed in Fig. 2.2. The quadrupole moment of the $12^+$ isomer is known, $Q_s = 0.65(5)$ $e$b [13]. Also the magnetic moments of both isomers are known [13, 14]. This provides the possibility for an internal calibration and allows us to deduce the quadrupole moment of the $11^-$ isomer from the ratio of the measured frequencies:

$$Q_s(11^-) = \frac{\nu_Q(11^-)}{\nu_Q(12^+)} \cdot Q_s(12^+) = (-)3.41 \pm 0.66 eb. \qquad (2)$$

The negative sign comes from systematics. The quoted uncertainty includes the uncertainties of the magnetic moments as well. Assuming an axial symmetry for this state (K=I=11), the quadrupole deformation is $\beta = (-)0.16(3)$, which is in a perfect agreement with the theory [15].

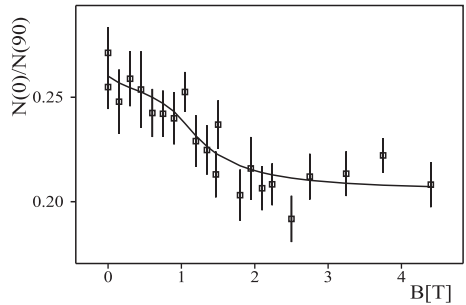

Fig. 2. LEMS curve for the $I^\pi = 11^-$ isomer in $^{196}$Pb. The recoiling nuclei were implanted into a Re polycrystalline foil at room temperature.

The $I^\pi = 11^-$ and the $I^\pi = 12^+$ isomers have the $\pi(s_{1/2}^{-2}h_{9/2}i_{13/2})$ and the $\nu(i_{13/2}^{-2})$ configurations, respectively, and the $\pi(s_{1/2}^{-2}h_{9/2}i_{13/2})_{11^-} \otimes \nu(i_{13/2}^{-2})_{12^+}$ 2p-2h configuration is suggested for the $I^\pi = 16^{(-)}$ magnetic bandhead [2]. It is possible to derive the quadrupole moment of this state, $Q_s(16^-) = -0.32(10)$ $eb$, using the additivity of the quadrupole operator, $Q_2^0 = Q_2^0(\pi) + Q_2^0(\nu)$:

$$Q_s(16^-) = <16,16|Q_2^0(\pi) + Q_2^0(\nu)|16,16>. \qquad (3)$$

This value proves the small deformation of the $16^{(-)}$ state, which supports the concept of magnetic rotation.

## 3. Conclusions

We have measured the spectroscopic quadrupole moments of high-spin isomers in $^{179}$W and $^{196}$Pb using the LEMS technique. These experiments were performed to test the predictions of the TAC model [3, 4]. We find that the deformation of the $K = \frac{35}{2}$ isomer in $^{179}$W is smaller than the ground-state deformation, which questions the current understanding of these excitations. For the $I^\pi = 16^{(-)}$ magnetic bandhead in $^{196}$Pb we derive a small quadrupole moment, which is in agreement with the concept of magnetic rotation.

Acknowledgements: GN is a post-doctoral researcher and DB and WDC are assistant researchers of FWO-Flanderen. DLB acknowledges grants from DWTC-Belgium and NATO. The work of the Bonn group is supported by BMFT (Germany) under contract No.06 BN907.